# Observation of the nonlinear chiral thermoelectric Hall effect in tellurium

Tetsuya Nomoto[1]*, Akiko Kikkawa[1], Kazuki Nakazawa[1], Terufumi Yamaguchi[1], Fumitaka Kagawa[1, 2]

[1] *RIKEN Center for Emergent Matter Science (CEMS), Wako 351-0198, Japan*

[2] *Department of Physics, Institute of Science Tokyo, Meguro 152-8551, Japan*

**Abstract**

The nonlinear thermoelectric effect is a key factor for realising unconventional thermoelectric phenomena, such as heat rectification and power generation using thermal fluctuations. Recent theoretical advances have indicated that chiral materials can host a variety of exotic nonlinear thermoelectric transport arising from inversion-symmetry breaking. However, experimental demonstration has yet to be achieved. Here, we report the observation of the nonlinear chiral thermoelectric Hall effect in chiral tellurium at room temperature, where a voltage is generated as a cross product of the temperature gradient and electric field. The resulting thermoelectric Hall voltage is on the order of μV, consistent with the prediction from the *ab initio* calculation. Furthermore, the sign of the thermoelectric Hall voltage is reversed depending on the crystal chirality, demonstrating a novel functionality of sign control of the thermoelectric effect by the chirality degrees of freedom. Our findings reveal the potential of chiral systems as nonlinear thermoelectric materials for advanced thermal management and energy harvesting.

**Main**

Chiral materials refer to systems with a noncentrosymmetric crystal structure that is not superimposable on its mirror image. The unique properties and phenomena exhibited by chiral materials have attracted great interest for a long time from both fundamental and applied perspectives[1–3]. Even in recent years, many intriguing phenomena have been discovered, such as the chirality-induced spin selectivity (CISS) effect[4–8], chiral phonons[9–11], and orbital angular momentum monopoles [12,13], all of which are expected to be applied to novel spintronics and optronics. Chiral materials are also interesting platforms for observing quantum transport phenomena caused by inversion symmetry breaking, which reflects the topological properties of the conduction electrons[14,15]. Recently, theoretical investigations have proposed that chiral systems can host a novel type of nonlinear thermoelectric effect of quantum origin arising from inversion symmetry breaking, which generates a voltage or temperature gradient as a higher-order response to the input electric current or heat flow[16–20]. Because this nonlinear thermoelectric effect reflects the quantum geometry of the electronic band structure, such as a Berry curvature dipole (BCD)[21], this effect may be referred to as the 'topological nonlinear thermoelectric (TNT) effect.' Fig. 1 shows various types of nonlinear thermoelectric effects that have been theoretically predicted thus far; these effects include the nonlinear Nernst effects[19,20], nonlinear Ettingshausen effects[17], nonlinear thermal Hall effects[18], and the nonlinear chiral thermoelectric (NCTE) Hall effect[22]. These effects are expected to provide an important basis for the development of advanced thermal management technologies, such as energy harvesting using thermal fluctuations[23] or heat rectification[24,25]. Furthermore, nonlinear thermoelectric effects are potentially sensitive probes for the topological properties of matter. Despite numerous theoretical predictions, however, these TNT phenomena have not yet been confirmed in real materials.

In this study, we experimentally demonstrate the NCTE Hall effect, as illustrated in Fig. 1d. The NCTE Hall effect is a phenomenon in which a current density $J_{\text{NCTE}}$ arises in the direction of the cross product of the electric field $E$ and the temperature gradient $\nabla T$ in chiral materials, which is described by the following:

$$J_{\text{NCTE}} = \sigma_{\text{NCTE}} \left( E \times \frac{-\nabla T}{T} \right), \tag{1}$$

where $\sigma_{\text{NCTE}}$ is the NCTE Hall conductivity and $T$ is the average temperature. The NCTE Hall effect has been theoretically predicted by chiral kinetic theory[26], hydrodynamics[27], semiclassical Boltzmann transport theory[16], and nonequilibrium Green's function method[22]. In contrast to the ordinary and anomalous Hall effects, the NCTE Hall effect can be observed even under time-reversal symmetry; specifically, an external magnetic field or internal magnetism is not needed for this effect. Furthermore, the symmetry analysis concludes that a finite NCTE Hall conductivity is allowed only for materials belonging to a chiral system (for a detailed discussion of the constraints arising from crystal symmetry, see Supplementary Material Section 1). Thus, a paramagnetic chiral system is a candidate for exploring the NCTE Hall effect.

**Characteristics of single crystal tellurium and the experimental setup**

In this work, we demonstrate the NCTE Hall effect using a typical chiral semiconductor, elemental tellurium (Te). Te has a crystal structure with helical chains extending along the *c*-axis, as shown in Fig. 2a and 2b, leading to two enantiomers of Te with different helix windings: right-handed (space group: $P3_121$, $D_3^4$) crystals and left-handed ($P3_221$, $D_3^6$) crystals. Irrespective of their chirality, all crystals behave as semiconductors with a small bandgap ($E_g \sim 0.34$ eV) when the impact of disorder is minimal[28]. Owing to the 'nested' band structure near the *H* point of the Brillouin zone, a large thermoelectric response is obtained even at room temperature[29,30]. Furthermore, a large thermal gradient is experimentally feasible for Te because of its low thermal conductivity ($\kappa \approx 2$ W K$^{-1}$ m$^{-1}$), which is mainly because of the complex crystal structure[29,31]. Thus, Te is likely an ideal platform for exploring phenomena driven by thermal gradients.

Fig. 2c and 2d shows our measurement setup for the NCTE Hall effect. To achieve a thermal gradient while maintaining the average temperature of the sample, we constructed a thermoelectric Hall effect measurement device that is combined with the seesaw heating method[32,33]. A single crystal of Te in a plate-like form was sandwiched by two thermometers, and two gold wires for current application and three gold wires for voltage detection were attached to the side surface. The chip heaters attached to the thermometers enabled a top-to-bottom or bottom-to-top temperature gradient in the sample. We applied a static temperature gradient and an AC excitation current with a frequency of 117.777 Hz perpendicular to the sample and measured the resulting Hall voltage along the direction of the cross product of the temperature gradient and current, and a standard five-terminal configuration combined with a lock-in technique was used; thus, the possibility of a signal due to the Seebeck effect being mixed in with the measured lock-in signal was negligible. More details are provided in the Methods section. We selected the $c$-axis as the direction of the Hall voltage detection and performed measurements under zero magnetic field. In this configuration with time-reversal symmetry, the $c$-axis Hall voltage due to the other second-order nonlinear effects [i.e., proportional to $E^2$, and $(\nabla T)^2$] is forbidden by symmetry (Supplementary Materials Sect. 1). Therefore, the present configuration is ideal for the detection of the NCTE Hall voltage[34].

**Observation of the NCTE Hall effect**

First, we checked the temperature gradient implemented with our experimental setup. Fig. 3a shows the time profiles of the top and bottom temperatures, $T_1$ and $T_2$, and the average temperature, $T_{\text{ave}} = (T_1 + T_2)/2$. For clarity, the resulting temperature difference ($\Delta T$) between the sample top and bottom is displayed in Fig. 3b. We define $\Delta T$ as positive when a temperature gradient is formed from top to bottom along the [1 0 0] direction of Te ($\Delta T = T_1 - T_2$). In our setup, a $\Delta T$ of $\approx \pm 10$ K over a 0.2 mm sample thickness (i.e., $\nabla T \approx \pm 1.7 \times 10^4$ K m$^{-1}$) was implemented, and the average temperature was kept constant with an accuracy of $\approx 0.1$ K.

The time profile of the resulting Hall voltage ($V_\text{H}$) is shown in Fig. 3c. The Hall voltage was clearly observed, and its sign changed in response to the reversal of the temperature gradient. To further confirm that the observed Hall voltage is caused by the coupling between the temperature gradient and the current, we investigated the temperature gradient and applied current dependences of the observed signal, and the results are shown in Fig. 3d and Fig. 3e, respectively. The observed Hall signal is proportional to both the temperature gradient and the applied current, which in good agreement with the characteristics of the NCTE Hall effect shown by Eq. (1). Furthermore, the observed linearity indicates that the Hall signal is not of extrinsic origin, such as Joule heating. To verify that this phenomenon is specific to single crystal tellurium and not due to experimental artefacts, we performed a similar experiment using the polycrystal bismuth (Bi); Bi is a nonchiral material, and its thermoelectric properties are similar to those of Te. As shown in Fig. 3f, a Hall voltage comparable to that observed for Te is absent for Bi, indicating that the Hall signal observed for Te is an intrinsic physical property of Te. Therefore, the observed Hall voltage in Te under the simultaneous application of current and a temperature gradient is consistent with the expectation for the NCTE Hall effect.

**Sign change upon reversing the crystal chirality**

Next, we show that the sign of the NCTE Hall signal is inverted in the two enantiomers of Te crystals with different chiralities as shown in Fig. 4a. To this end, we perform similar measurements using sample B, which has the opposite chirality to sample A. According to the etch pit method[35], sample A, which has 4-shaped etch pits (Fig. 4b), is identified as a right-handed crystal, whereas sample B, which has inverse-4-shaped etch pits (Fig. 4c), is identified as a left-handed crystal. Both samples A and B exhibit a semiconducting temperature dependence of the resistance near room temperature and the same sign of the normal Hall coefficient (see Supplementary Materials Section 3). The NCTE Hall effect measurements for each crystal are shown in Figs. 4d and 4e, and its variation with the temperature gradient is plotted in Fig. 4f. Both crystals exhibit linear NCTE Hall signals as a function of the temperature gradient, and these crystals have similar absolute values but opposite signs. Theoretical considerations suggest that the sign of the NCTE Hall effect in semiconducting Te is insensitive to changes in both the type of carrier (see Supplementary Materials Section 4) and the position of the chemical potential due to self-doping[34]. Therefore, the observed sign reversal is attributed to the chirality inversion. Furthermore, for Eq. (1) to hold under spatial inversion and mirror operation, the sign of $\sigma_{\mathrm{NCTE}}$ should be inverted for the two enantiomers of the Te crystals with different chiralities and thus our observation is in agreement with Eq. (1). This result highlights an intriguing aspect of the TNT effect in chiral materials; the sign of the thermoelectric response is governed by the chirality degree of freedom. This finding is highly important since it represents a unique feature that is absent in conventional thermoelectric effects.

**Comparison with the theoretical prediction**

Finally, we compare our experimental results with current theoretical predictions for the NCTE Hall effect. The microscopic theory[22] suggests that the NCTE Hall effect can be described as the sum of two contributions: one from the BCD and the other from the orbital magnetic moment (OM) (see Supplementary Materials Section 4 for details). Under the assumption that the relaxation time approximation holds, DFT calculations[34] show that the NCTE Hall current density in sample A is $j_{\mathrm{NCTE}}$ = 6.0–7.2 × $10^{-4}$ A m$^{-2}$ (see Methods for calculation details), whereas calculating $j_{\mathrm{NCTE}}$ from the observed NCTE Hall voltage results in 3.3 × $10^{-3}$ A m$^{-2}$. The theoretical value is approximately one-fifth of the experimental value, indicating that both values agree within a factor of five. Therefore, the current microscopic theoretical estimates appear to effectively reproduce our experimental results. Nevertheless, we consider this comparison preliminary. On the experimental side, our order-of-magnitude estimate is based on the carrier density and effective mass obtained using the single-carrier approximation, whereas the application of a multicarrier model does not necessarily produce reasonable results. On the theoretical side, the current DFT calculations are performed under the relaxation time approximation and do not take into account the energy dependence of the scattering time.

**Discussion**

The NCTE Hall effect confirmed in this experiment is a new transport phenomenon, and several future research directions are discussed below. First, the NCTE Hall effect may be viewed as a phenomenon in which the electron flow is bent by an effective magnetic field generated by a temperature gradient. When we use the Hall coefficient of the ordinary Hall effect of sample A to convert it into an equivalent effective magnetic field, we find it to be $\approx$ 4.4 mT for $\nabla T/T = 61.1$ m$^{-1}$. In addition to exploring how to enhance the NCTE Hall effect, the impact of the temperature gradient-induced effective magnetic field on other physical properties is a potential future research direction. Second, the NCTE Hall effect may be a useful probe of BCD and OM. The former is an important property for device applications of chiral materials because it is the origin of the peculiar nonlinear transport phenomena such as the nonlinear Hall effect in chiral systems [14]. The latter is likely important for the development of orbital electronics since it can aid in the elucidation of the unique orbital magnetic texture and its associated phenomena, such as orbital magnetic monopoles[12,13,36], the orbital Edelstein effect[37], and the anomalous Hall effect[38]. Finally, we emphasize that various nonlinear thermoelectric effects are predicted with the framework of the semiclassical Boltzmann transport equation, as is the case for the NCTE Hall effect [16]. The discovery of the NCTE Hall effect spurs the expectation that other nonlinear thermoelectric phenomena should also be observed in real materials. Exploration of the nonlinear thermoelectric phenomena, especially with a focus on noncentrosymmetric materials, is important from a fundamental point of view and can likely aid in the development of exotic thermal control and thermoelectric devices.

**Method**

**Sample Preparation**

Single crystals of trigonal Te were grown using the physical vapour transport (PVT) method[39]. For the PVT method, 0.36 g of elemental Te (6N) grains were sealed in an evacuated quartz tube and heated in a three-zone tube furnace with temperatures set to 420°C (source side) and 330°C (growth side). The details of the synthesis process are described in Supplementary Materials Section 1. The single crystals were columnar, with a typical length of approximately 1 cm and a width of 2–4 mm. The samples were cut with a diamond cutter and a wire saw. The chirality of the crystal was determined by observing the shapes of the etch pits produced by the slow action of hot sulfuric acid (100°C, 30 min) on the cleavage planes of the crystals[35].

**NCTE Hall measurement**

The Hall measurement was performed by the standard five-terminal method[40–42]. In this method, by adjusting the equipotential surface in the crystals using a potentiometer, the contamination of the longitudinal resistive voltage to the Hall electrodes can be eliminated. These zero adjustments were performed under the condition of $\Delta T = 0$. Five gold wires ($\varphi = 10$ μm) were attached to the (1 0 -1 0) surface, and the electrodes for detecting the Hall voltage were parallel to the [001] direction to eliminate the other contributions of nonlinear transport (details in Supplementary Materials Sect. 1). The NCTE Hall voltage was detected using a lock-in amplifier (DSP 7240, Signal Recovery Co.) and a low-noise preamplifier (SR560, Stanford Research Systems). A temperature gradient perpendicular to the current was formed using a $RuO_2$ chip heater (1 kΩ). The magnitude of the temperature gradient was obtained by two Pt thin film (PT-100) thermometers. During the measurement, two AC/DC current sources (Model 6221, Keithley Instrument) were used to independently control the temperatures above and below the sample to adjust the average temperature and temperature gradient of the sample. The chip heaters, sample, and thermometers were thermally connected by G10 varnish. The current value flowing through the sample was determined by sensing the current through the shunt resistor with another lock-in amplifier. The frequency $f$ of the AC power supply was set to 117.777 Hz. The Hall current density $j_{NCTE}$ used for comparison with theory was obtained by dividing the observed Hall voltage by the resistance of the measurement circuit.

**Quantitative comparison with DFT calculations**

We calculated the expected value of the NCTE Hall current by referring to the *ab initio* calculation in Ref. 34. A tabular representation of the NCTE Hall effect, on which the calculations are based, is presented in Supplementary Materials Section 4. We assumed that the chemical potential, $\mu$, of the measured sample was in the in-gap region; specifically, $\mu$ ranged from -0.08 eV to 0.05 eV, which is a reasonable setting since the temperature dependence of the resistivity of the sample shows semiconducting behaviour (see Supplementary Materials Section 3). Furthermore, since the experiments were performed at room temperature, the temperature parameter $k_B T = 0.03$ eV was used for this calculation, where $k_B$ is the Boltzmann constant, since the experiments were performed at room temperature. Here, for sample A at 300 K, the longitudinal resistivity at zero magnetic field and Hall coefficient in a low magnetic field ($|B| < 1$ T) are $\rho_{xx} = 1.08 \times 10^{-3}$ Ω m and $R_H = -6.88 \times 10^{-5}$ m$^3$C$^{-1}$, respectively (see Supplementary Materials Section 3). The density of the electron carriers determined from the $R_H$ is $n = 9.07 \times 10^{22}$ m$^{-3}$. The effective mass of the electrons $m^* = 0.45 m_0$ is taken from the literature[43], where $m_0$ is the free electron mass ($m_0 = 9.11 \times 10^{-31}$ kg). Using the values of $\rho$, $n$, $m^*$, and the elementary charge $e$ of $1.60 \times 10^{-19}$ C, the relaxation time $\tau$ is calculated using the Drude model ($\tau = m^*/ne^2\rho$) to be approximately 163 fs. The external fields used in the present experiments are as follows: $E = 1.80$ V m$^{-1}$ and $\nabla T/T = 61.1$ m$^{-1}$. Assuming that the NCTE Hall current is linearly proportional to $\tau$, the theoretical value of the NCTE Hall current density $j_{NCTE}$ under the present experimental conditions is $6.0$–$7.2 \times 10^{-4}$ A m$^{-2}$. For the same external field parameters, the observed value is $V_H = 0.61$ μV, and the Hall current is determined to be $3.3 \times 10^{-3}$ A m$^{-2}$ by using the dimensions of sample A ($l \times w \times t = 0.90$ mm $\times$ 1.1 mm $\times$ 0.60 mm), and the circuit resistance for the Hall measurement ($R = 283$ Ω).

## Acknowledgements
We would like to thank Tetsuya Furukawa for advice on sample preparation and Takahisa Arima for fruitful discussions. This research was supported by the Japan Society for the Promotion of Science (JSPS) KAKENHI (Grant Nos. 21K13875, 21K20347 and 23K13054).

## Author information

### Contributions
T.N. and F.K. conceived the project. T.N. and A.K. performed sample growth by PVT, X-ray (Laue) measurements, and sample cutting. T.N. constructed and evaluated the measurement setup for the NCTE Hall effect. T.N. analysed the obtained data. T.N., K.N., T.Y., and F.K. were responsible for discussion of the symmetry and quantitative analysis. F.K. supervised this project. All authors contributed to the writing of the manuscript.

### Corresponding author
Correspondence should be addressed to Tetsuya Nomoto.

## Ethics declarations
### Competing interests
The authors declare that they have no competing interests.

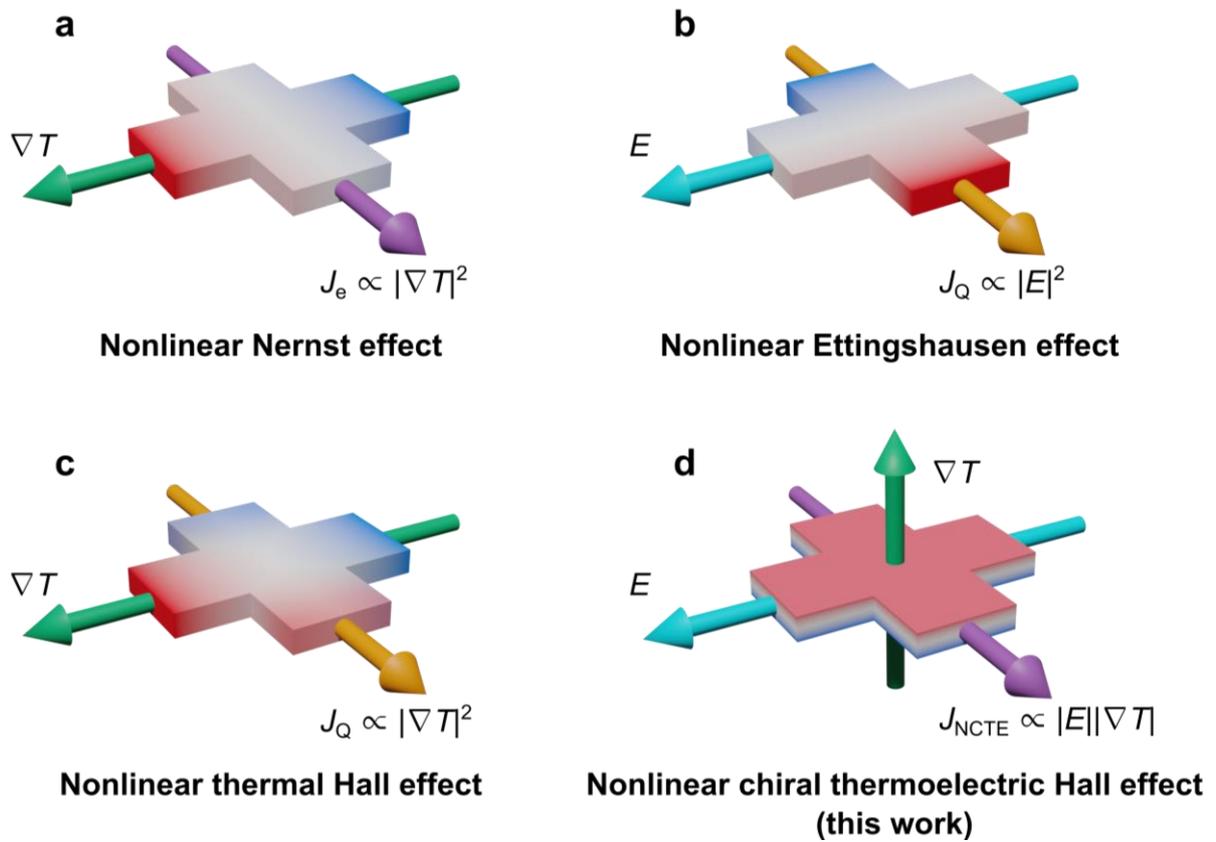

**Figure 1| Nonlinear thermoelectric effects in inversion symmetry-broken systems. a**, Nonlinear Nernst effect. A transverse current flow $J_e$ proportional to the square of a temperature gradient, $|\nabla T|^2$, is generated. **b**, Nonlinear Ettingshausen effect. A transverse heat flow $J_Q$ proportional to the square of an electric field, $|E|^2$, is generated. **c**, Nonlinear thermal Hall effect. The $J_Q$ proportional to the square of an electric field, $|\nabla T|^2$, is generated. **d**, Nonlinear chiral-thermoelectric (NCTE) Hall effect investigated in this work. The NCTE Hall current $J_{\text{NCTE}}$ is generated as the cross product of $E$ and $\nabla T$.

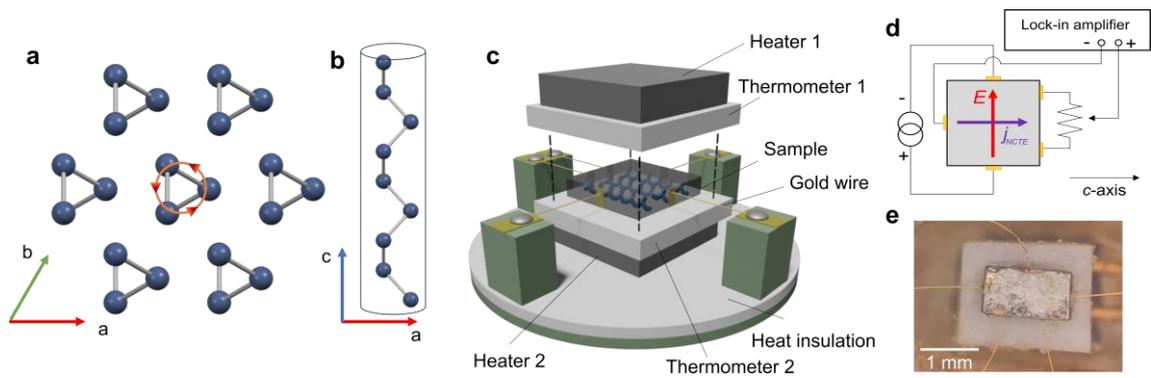

**Figure 2| Experimental setup for the NCTE Hall effect. a**, Crystal structure of Te with space group P3$_1$21 (right-handed crystal). **b**, Side view of Te. The atomic chain is along the *c*-axis. **c**, Schematic illustration of the NCTE Hall effect measurement system. The target sample is sandwiched by two thermometers with a heater. A temperature gradient is formed in the sample along the *a*-axis, and its direction can be reversed. Five gold wires are attached to the side of the sample to measure the Hall effect. **d**, Schematic diagram of the circuit. Both the AC excitation current and the temperature gradient are applied perpendicular to the *c*-axis, resulting in the NCTE Hall voltage along the *c*-axis. **e**, Picture of the measured sample.

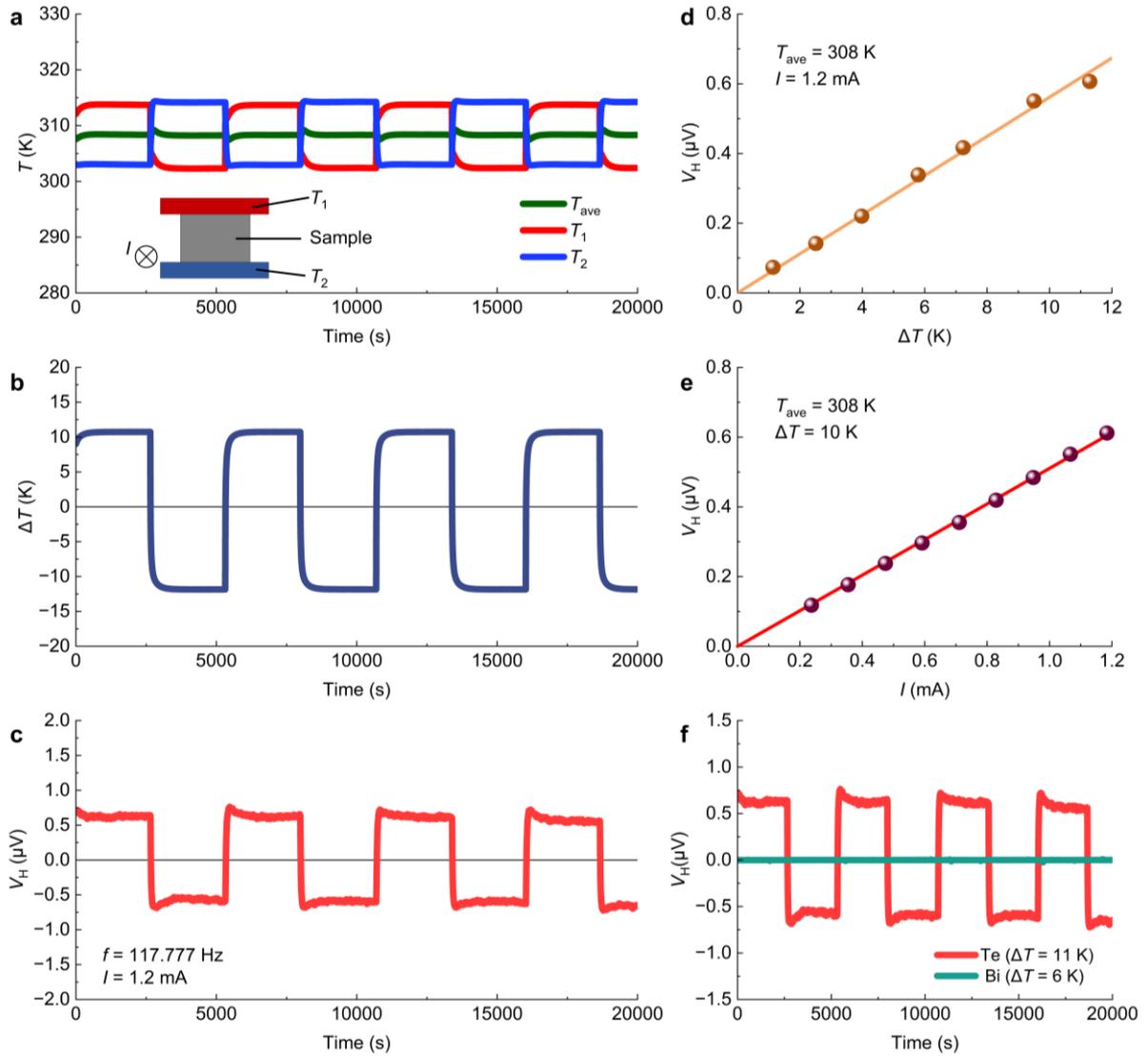

**Figure 3| Observation of the NCTE Hall effect. a**, Time profiles of temperatures at the top ($T_1$) and bottom ($T_2$) of the sample and the average temperature ($T_{ave}$). $T_{ave}$ in the temperature stability region is in the range of approximately ±0.1 K. **b**, Time profile of the temperature gradient ($\Delta T$) in the sample. **c**, Signal of the observed NCTE Hall effect ($V_H$). The Hall voltage oscillates in phase with the oscillation of $\Delta T$. **d**, $\Delta T$ versus $V_H$ plot. **e**, $I$ versus $V_H$ plot. The errors of the standard deviation are within the points. The observed Hall voltages are proportional to both $\Delta T$ and $I$, which is consistent with the predicted feature of the NCTE Hall effect. **f**, Comparison between Te single crystal (red line) and Bi polycrystal (green line). No Hall voltage corresponding to the amplitude of the temperature gradient was observed in Bi.

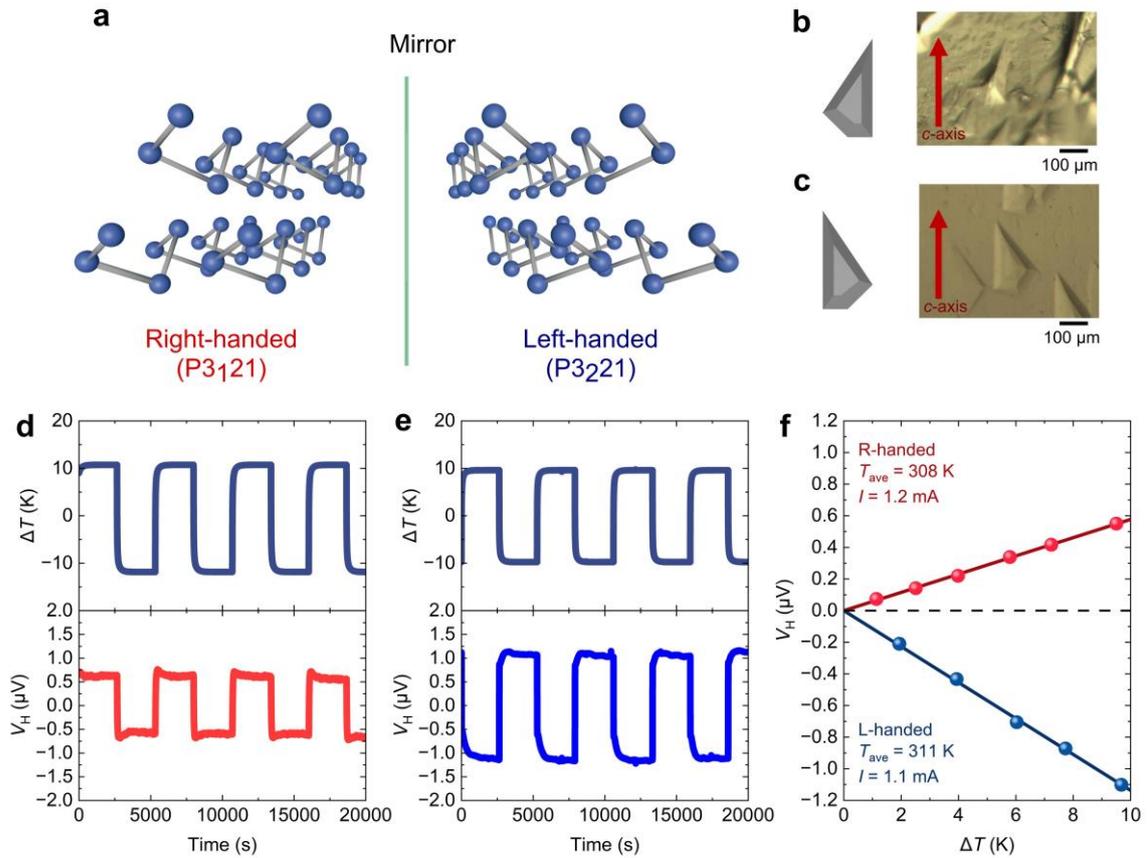

**Figure 4| Chirality dependence of the NCTE Hall effect. a**, Two enantiomers of Te, right-handed (space group: P3$_1$21) and left-handed (space group: P3$_2$21) crystals. **b**, 4-shaped etch pits appearing on the surface of sample A. **c**, Inverse-4-shaped etch pits appearing on the surface of sample B. The 4-shaped and inverse-4-shaped etch pits indicate right-handed and left-handed crystals, respectively. **d**, NCTE Hall effect of sample A. **e**, NCTE Hall effect of sample B. The sign of the NCTE Hall voltage produced by the temperature gradient is reversed depending on the chirality of the crystals. **f**, $\Delta T$ dependence of the NCTE Hall effect in samples A and B. The errors of the standard deviation are within the points.

**Supplementary Materials for**

**"Observation of the nonlinear chiral thermoelectric Hall effect in tellurium"**

Tetsuya Nomoto[1]*, Akiko Kikkawa[1], Kazuki Nakazawa[1], Terufumi Yamaguchi[1], Fumitaka Kagawa[1,2]

[1] *RIKEN Center for Emergent Matter Science (CEMS), Wako 351-0198, Japan*
[2] *Department of Physics, Institute of Science Tokyo, Meguro 152-8551, Japan*

§1 **Symmetry constraints on the NCTE Hall effect**
§2 **Sample preparation**
§3 **Electrical properties of the tellurium samples**
§4 **Microscopic theory of the NCTE Hall effect**

## §1 Symmetry constraints on the NCTE Hall effect

In this section, we explain the conditions under which the NCTE Hall effect is observed based on the symmetry analysis. The electric current $\boldsymbol{J}$ driven by both an electric field and a temperature gradient is denoted as follows:

$$J_i = \sigma_{ijk} E_j \partial_k \left(-\frac{\partial_k T}{T}\right), \tag{S1}$$

where $\sigma_{ijk}$ is the third-rank nonlinear thermoelectric conductivity tensor with $i, j, k \in \{1, 2, 3\}$. The indices 1, 2, and 3 are associated with the cartesian coordinates $x$, $y$, and $z$, respectively. The component tables of the third-rank conductivity tensor $\chi$ for two different external fields in chiral crystals are provided in Tables S1. Since tellurium (Te) belongs to the point group 32, the nonlinear thermoelectric conductivity tensor in Te is expressed as follows[1]:

$$\boldsymbol{\sigma} = \begin{pmatrix} \sigma_{xxx} & 0 & 0 & 0 & \sigma_{xyy} & \sigma_{xzy} & 0 & \sigma_{xyz} & 0 \\ 0 & \sigma_{yyx} & \sigma_{yzx} & \sigma_{yxy} & 0 & 0 & \sigma_{yxz} & 0 & 0 \\ 0 & 0 & \sigma_{zyx} & 0 & \sigma_{zxy} & 0 & 0 & 0 & 0 \end{pmatrix} \tag{S2}$$

where $\sigma_{xxx} = -\sigma_{xyy} = -\sigma_{yyx} = -\sigma_{yxy}$, $\sigma_{xzy} = -\sigma_{yzx}$, $\sigma_{xyz} = -\sigma_{yxz}$, and $\sigma_{zxy} = -\sigma_{zyx}$. The coordinates $x$ and $z$ correspond to the $a$-axis and $c$-axis, respectively. Here, the NCTE Hall current $\boldsymbol{J} = (J_x, J_y, J_z)$ is given by the following:

$$J_x = \frac{\sigma_{xzy} - \sigma_{xyz}}{2}\left(E_z\left(-\frac{\partial_y T}{T}\right) - E_y\left(-\frac{\partial_z T}{T}\right)\right), \tag{S3}$$

$$J_y = \frac{\sigma_{yzx} - \sigma_{yxz}}{2}\left(E_z\left(-\frac{\partial_x T}{T}\right) - E_x\left(-\frac{\partial_z T}{T}\right)\right), \tag{S4}$$

$$J_z = \frac{\sigma_{zyx} - \sigma_{zxy}}{2}\left(E_y\left(-\frac{\partial_x T}{T}\right) - E_x\left(-\frac{\partial_y T}{T}\right)\right). \tag{S5}$$

Therefore, the NCTE Hall effect in the $x$, $y$, and $z$ directions in Te is finite. Similarly, for other chiral point groups, the NCTE Hall effect can have a finite value based on symmetry analysis. On the other hand, for noncentrosymmetric but achiral cases, the NCTE Hall current is zero when it is calculated using Tables S2. Therefore, the system must be chiral to observe the NCTE Hall effect. Note that when the contribution from the Berry curvature dipole (BCD) in the NCTE Hall effect is considered, an additional constraint is added: the BCD tensor needs to be traceless[2]. In the case of Te, the BCD contribution is not forbidden because of the crystal symmetry[3, 4].

Considering the second-order nonlinear current response due to electric fields, $J'_i = \sigma'_{ijk} E_j E_k$, the currents generated in the $x$, $y$, and $z$ directions in the systems with point group 32 are as follows:

$$J'_x = \sigma'_{xxx}(E_x^2 - E_y^2) + (\sigma'_{xzy} + \sigma'_{xyz})E_y E_z, \tag{S6}$$

$$J'_y = 2\sigma'_{yxy} E_x E_y + (\sigma'_{yxz} + \sigma'_{yxz})E_x E_z, \tag{S7}$$

$$J'_z = 0. \tag{S8}$$

In the *c*-axis direction, the second-order nonlinear electrical response is forbidden by symmetry[3]. This symmetry constraint also applies to the second-order response of the temperature gradient $(\nabla T)^2$. Therefore, the NCTE Hall currents in the *c*-axis direction of Te are unaffected by both the second-order nonlinear responses and the linear response.

**Table S1 | Components of the third-rank conductivity tensor $\chi_{ijk}$ for the chiral point groups.** The red cells in the tables represent finite components.

**1**

| $\chi_{ijk}$ | | jk | | | | | | | | |
|---|---|---|---|---|---|---|---|---|---|---|
| | | 11 | 21 | 31 | 12 | 22 | 32 | 13 | 23 | 33 |
| i | 1 | $\chi_{111}$ | $\chi_{121}$ | $\chi_{131}$ | $\chi_{112}$ | $\chi_{122}$ | $\chi_{132}$ | $\chi_{113}$ | $\chi_{123}$ | $\chi_{133}$ |
| | 2 | $\chi_{211}$ | $\chi_{221}$ | $\chi_{231}$ | $\chi_{212}$ | $\chi_{222}$ | $\chi_{232}$ | $\chi_{213}$ | $\chi_{223}$ | $\chi_{233}$ |
| | 3 | $\chi_{311}$ | $\chi_{321}$ | $\chi_{331}$ | $\chi_{312}$ | $\chi_{322}$ | $\chi_{332}$ | $\chi_{313}$ | $\chi_{323}$ | $\chi_{333}$ |

Number of independent coefficients: 27

**2**

| $\chi_{ijk}$ | | jk | | | | | | | | |
|---|---|---|---|---|---|---|---|---|---|---|
| | | 11 | 21 | 31 | 12 | 22 | 32 | 13 | 23 | 33 |
| i | 1 | 0 | $\chi_{121}$ | 0 | $\chi_{112}$ | 0 | $\chi_{132}$ | 0 | $\chi_{123}$ | 0 |
| | 2 | $\chi_{211}$ | 0 | $\chi_{231}$ | 0 | $\chi_{222}$ | 0 | $\chi_{213}$ | 0 | $\chi_{233}$ |
| | 3 | 0 | $\chi_{321}$ | 0 | $\chi_{312}$ | 0 | $\chi_{332}$ | 0 | $\chi_{323}$ | 0 |

Number of independent coefficients: 13

**222**

| $\chi_{ijk}$ | | jk | | | | | | | | |
|---|---|---|---|---|---|---|---|---|---|---|
| | | 11 | 21 | 31 | 12 | 22 | 32 | 13 | 23 | 33 |
| i | 1 | 0 | 0 | 0 | 0 | 0 | $\chi_{132}$ | 0 | $\chi_{123}$ | 0 |
| | 2 | 0 | 0 | $\chi_{231}$ | 0 | 0 | 0 | $\chi_{213}$ | 0 | 0 |
| | 3 | 0 | $\chi_{321}$ | 0 | $\chi_{312}$ | 0 | 0 | 0 | 0 | 0 |

Number of independent coefficients: 6

**23**

| $\chi_{ijk}$ | | jk | | | | | | | | |
|---|---|---|---|---|---|---|---|---|---|---|
| | | 11 | 21 | 31 | 12 | 22 | 32 | 13 | 23 | 33 |
| i | 1 | 0 | 0 | 0 | 0 | 0 | $\chi_{132}$ | 0 | $\chi_{123}$ | 0 |
| | 2 | 0 | 0 | $\chi_{231}$ | 0 | 0 | 0 | $\chi_{213}$ | 0 | 0 |
| | 3 | 0 | $\chi_{321}$ | 0 | $\chi_{312}$ | 0 | 0 | 0 | 0 | 0 |

Number of independent coefficients: 2

$\chi_{132} = \chi_{213} = \chi_{321}$, $\chi_{123} = \chi_{231} = \chi_{312}$

**3**

| $\chi_{ijk}$ | | jk | | | | | | | | |
|---|---|---|---|---|---|---|---|---|---|---|
| | | 11 | 21 | 31 | 12 | 22 | 32 | 13 | 23 | 33 |
| i | 1 | $\chi_{111}$ | $\chi_{121}$ | $\chi_{131}$ | $\chi_{112}$ | $\chi_{122}$ | $\chi_{132}$ | $\chi_{113}$ | $\chi_{123}$ | 0 |
| | 2 | $\chi_{211}$ | $\chi_{221}$ | $\chi_{231}$ | $\chi_{212}$ | $\chi_{222}$ | $\chi_{232}$ | $\chi_{213}$ | $\chi_{223}$ | 0 |
| | 3 | $\chi_{311}$ | $\chi_{321}$ | 0 | $\chi_{312}$ | $\chi_{322}$ | 0 | 0 | 0 | $\chi_{333}$ |

Number of independent coefficients: 9

$\chi_{111} = -\chi_{122} = -\chi_{221} = -\chi_{212}$, $\chi_{112} = \chi_{121} = \chi_{211} = -\chi_{222}$, $\chi_{131} = \chi_{232}$, $\chi_{132} = -\chi_{231}$

$\chi_{113} = \chi_{223}$, $\chi_{123} = -\chi_{213}$, $\chi_{311} = \chi_{322}$, $\chi_{312} = -\chi_{321}$

**Table S1 (continued)**

| $\chi_{ijk}$ | | 32 | | | | | | | | |
|---|---|---|---|---|---|---|---|---|---|---|
| | | \multicolumn{9}{c|}{$jk$} |
| | | 11 | 21 | 31 | 12 | 22 | 32 | 13 | 23 | 33 |
| $i$ | 1 | $\chi_{111}$ | 0 | 0 | 0 | $\chi_{122}$ | $\chi_{132}$ | 0 | $\chi_{123}$ | 0 |
| | 2 | 0 | $\chi_{221}$ | $\chi_{231}$ | $\chi_{212}$ | 0 | 0 | $\chi_{213}$ | 0 | 0 |
| | 3 | 0 | $\chi_{321}$ | 0 | $\chi_{312}$ | 0 | 0 | 0 | 0 | 0 |

Number of independent coefficients: 4

$\chi_{111} = -\chi_{122} = -\chi_{221} = -\chi_{212}$, $\chi_{132} = -\chi_{231}$, $\chi_{123} = -\chi_{213}$, $\chi_{312} = -\chi_{321}$

| $\chi_{ijk}$ | | 4 | | | | | | | | |
|---|---|---|---|---|---|---|---|---|---|---|
| | | \multicolumn{9}{c|}{$jk$} |
| | | 11 | 21 | 31 | 12 | 22 | 32 | 13 | 23 | 33 |
| $i$ | 1 | 0 | 0 | $\chi_{131}$ | 0 | 0 | $\chi_{132}$ | $\chi_{113}$ | $\chi_{123}$ | 0 |
| | 2 | 0 | 0 | $\chi_{231}$ | 0 | 0 | $\chi_{232}$ | $\chi_{213}$ | $\chi_{223}$ | 0 |
| | 3 | $\chi_{311}$ | $\chi_{321}$ | 0 | $\chi_{312}$ | $\chi_{322}$ | 0 | 0 | 0 | $\chi_{333}$ |

Number of independent coefficients: 7

$\chi_{131} = -\chi_{232}$, $\chi_{132} = -\chi_{231}$, $\chi_{113} = \chi_{223}$, $\chi_{123} = -\chi_{213}$, $\chi_{311} = \chi_{322}$, $\chi_{312} = -\chi_{321}$

| $\chi_{ijk}$ | | 422 | | | | | | | | |
|---|---|---|---|---|---|---|---|---|---|---|
| | | \multicolumn{9}{c|}{$jk$} |
| | | 11 | 21 | 31 | 12 | 22 | 32 | 13 | 23 | 33 |
| $i$ | 1 | 0 | 0 | 0 | 0 | 0 | $\chi_{132}$ | 0 | $\chi_{123}$ | 0 |
| | 2 | 0 | 0 | $\chi_{231}$ | 0 | 0 | 0 | $\chi_{213}$ | 0 | 0 |
| | 3 | 0 | $\chi_{321}$ | 0 | $\chi_{312}$ | 0 | 0 | 0 | 0 | 0 |

Number of independent coefficients: 3

$\chi_{132} = -\chi_{231}$, $\chi_{123} = -\chi_{213}$, $\chi_{312} = -\chi_{321}$

| $\chi_{ijk}$ | | 432 | | | | | | | | |
|---|---|---|---|---|---|---|---|---|---|---|
| | | \multicolumn{9}{c|}{$jk$} |
| | | 11 | 21 | 31 | 12 | 22 | 32 | 13 | 23 | 33 |
| $i$ | 1 | 0 | 0 | 0 | 0 | 0 | $\chi_{132}$ | 0 | $\chi_{123}$ | 0 |
| | 2 | 0 | 0 | $\chi_{231}$ | 0 | 0 | 0 | $\chi_{213}$ | 0 | 0 |
| | 3 | 0 | $\chi_{321}$ | 0 | $\chi_{312}$ | 0 | 0 | 0 | 0 | 0 |

Number of independent coefficients: 1

$\chi_{123} = -\chi_{132} = \chi_{231} = -\chi_{213} = -\chi_{321} = \chi_{312}$

| $\chi_{ijk}$ | | 6 | | | | | | | | |
|---|---|---|---|---|---|---|---|---|---|---|
| | | \multicolumn{9}{c|}{$jk$} |
| | | 11 | 21 | 31 | 12 | 22 | 32 | 13 | 23 | 33 |
| $i$ | 1 | 0 | 0 | $\chi_{131}$ | 0 | 0 | $\chi_{132}$ | $\chi_{113}$ | $\chi_{123}$ | 0 |
| | 2 | 0 | 0 | $\chi_{231}$ | 0 | 0 | $\chi_{232}$ | $\chi_{213}$ | $\chi_{223}$ | 0 |
| | 3 | $\chi_{311}$ | $\chi_{321}$ | 0 | $\chi_{312}$ | $\chi_{322}$ | 0 | 0 | 0 | $\chi_{333}$ |

Number of independent coefficients: 7

$\chi_{131} = \chi_{232}$, $\chi_{132} = -\chi_{231}$, $\chi_{113} = \chi_{223}$, $\chi_{123} = -\chi_{213}$, $\chi_{311} = \chi_{322}$, $\chi_{312} = -\chi_{321}$

**Table S1 (continued)**

| $\chi_{ijk}$ | | 622 | | | | | | | | |
|---|---|---|---|---|---|---|---|---|---|---|
| | | *jk* | | | | | | | | |
| | | 11 | 21 | 31 | 12 | 22 | 32 | 13 | 23 | 33 |
| *i* | 1 | 0 | 0 | 0 | 0 | 0 | $\chi_{132}$ | 0 | $\chi_{123}$ | 0 |
| | 2 | 0 | 0 | $\chi_{231}$ | 0 | 0 | 0 | $\chi_{213}$ | 0 | 0 |
| | 3 | 0 | $\chi_{321}$ | 0 | $\chi_{312}$ | 0 | 0 | 0 | 0 | 0 |
| Number of independent coefficients: 3 | | | | | | | | | | |
| $\chi_{132} = -\chi_{231}$, $\chi_{123} = -\chi_{213}$, $\chi_{312} = -\chi_{321}$ | | | | | | | | | | |

**Table S2 | Components of the third-rank conductivity tensor $\chi_{ijk}$ for the achiral point groups.**
The red cells in the tables represent finite components.

### m

| $\chi_{ijk}$ | | \multicolumn{9}{c}{jk} |
|---|---|---|---|---|---|---|---|---|---|
| | | 11 | 21 | 31 | 12 | 22 | 32 | 13 | 23 | 33 |
| i | 1 | $\chi_{111}$ | 0 | $\chi_{131}$ | 0 | $\chi_{122}$ | 0 | $\chi_{113}$ | 0 | $\chi_{133}$ |
| | 2 | 0 | $\chi_{221}$ | 0 | $\chi_{212}$ | 0 | $\chi_{232}$ | 0 | $\chi_{223}$ | 0 |
| | 3 | $\chi_{311}$ | 0 | $\chi_{331}$ | 0 | $\chi_{322}$ | 0 | $\chi_{313}$ | 0 | $\chi_{333}$ |

Number of independent coefficients: 14

### mm2

| $\chi_{ijk}$ | | \multicolumn{9}{c}{jk} |
|---|---|---|---|---|---|---|---|---|---|
| | | 11 | 21 | 31 | 12 | 22 | 32 | 13 | 23 | 33 |
| i | 1 | 0 | 0 | $\chi_{131}$ | 0 | 0 | 0 | $\chi_{113}$ | 0 | 0 |
| | 2 | 0 | 0 | 0 | 0 | 0 | $\chi_{232}$ | 0 | $\chi_{223}$ | 0 |
| | 3 | $\chi_{311}$ | 0 | 0 | 0 | $\chi_{322}$ | 0 | 0 | 0 | $\chi_{333}$ |

Number of independent coefficients: 7

### 3m

| $\chi_{ijk}$ | | \multicolumn{9}{c}{jk} |
|---|---|---|---|---|---|---|---|---|---|
| | | 11 | 21 | 31 | 12 | 22 | 32 | 13 | 23 | 33 |
| i | 1 | 0 | $\chi_{121}$ | $\chi_{131}$ | $\chi_{112}$ | 0 | 0 | $\chi_{113}$ | 0 | 0 |
| | 2 | $\chi_{211}$ | 0 | 0 | 0 | $\chi_{222}$ | $\chi_{232}$ | 0 | $\chi_{223}$ | 0 |
| | 3 | $\chi_{311}$ | 0 | 0 | 0 | $\chi_{322}$ | 0 | 0 | 0 | $\chi_{333}$ |

Number of independent coefficients: 5

$\chi_{112} = \chi_{221} = \chi_{211} = -\chi_{222}$, $\chi_{131} = \chi_{232}$, $\chi_{113} = \chi_{223}$, $\chi_{311} = \chi_{322}$

### -4

| $\chi_{ijk}$ | | \multicolumn{9}{c}{jk} |
|---|---|---|---|---|---|---|---|---|---|
| | | 11 | 21 | 31 | 12 | 22 | 32 | 13 | 23 | 33 |
| i | 1 | 0 | 0 | $\chi_{131}$ | 0 | 0 | $\chi_{132}$ | $\chi_{113}$ | $\chi_{123}$ | 0 |
| | 2 | 0 | 0 | $\chi_{231}$ | 0 | 0 | $\chi_{232}$ | $\chi_{213}$ | $\chi_{223}$ | 0 |
| | 3 | $\chi_{311}$ | $\chi_{321}$ | 0 | $\chi_{312}$ | $\chi_{322}$ | 0 | 0 | 0 | 0 |

Number of independent coefficients: 6

$\chi_{132} = \chi_{231}$, $\chi_{131} = -\chi_{232}$, $\chi_{113} = -\chi_{223}$, $\chi_{123} = \chi_{213}$, $\chi_{311} = -\chi_{322}$, $\chi_{312} = \chi_{321}$

### -42m

| $\chi_{ijk}$ | | \multicolumn{9}{c}{jk} |
|---|---|---|---|---|---|---|---|---|---|
| | | 11 | 21 | 31 | 12 | 22 | 32 | 13 | 23 | 33 |
| i | 1 | 0 | 0 | 0 | 0 | 0 | $\chi_{132}$ | 0 | $\chi_{123}$ | 0 |
| | 2 | 0 | 0 | $\chi_{231}$ | 0 | 0 | 0 | $\chi_{213}$ | 0 | 0 |
| | 3 | 0 | $\chi_{321}$ | 0 | $\chi_{312}$ | 0 | 0 | 0 | 0 | 0 |

Number of independent coefficients: 3

$\chi_{132} = \chi_{231}$, $\chi_{123} = \chi_{213}$, $\chi_{312} = \chi_{321}$

**Table S2 (continued)**

### -43m

| $\chi_{ijk}$ | | \multicolumn{9}{c}{jk} | | | | | | | | |
|---|---|---|---|---|---|---|---|---|---|---|
| | | 11 | 21 | 31 | 12 | 22 | 32 | 13 | 23 | 33 |
| i | 1 | 0 | 0 | 0 | 0 | 0 | $\chi_{132}$ | 0 | $\chi_{123}$ | 0 |
| | 2 | 0 | 0 | $\chi_{231}$ | 0 | 0 | 0 | $\chi_{213}$ | 0 | 0 |
| | 3 | 0 | $\chi_{321}$ | 0 | $\chi_{312}$ | 0 | 0 | 0 | 0 | 0 |

Number of independent coefficients: 1

$\chi_{123} = \chi_{132} = \chi_{231} = \chi_{213} = \chi_{321} = \chi_{312}$

### 4mm

| $\chi_{ijk}$ | | \multicolumn{9}{c}{jk} | | | | | | | | |
|---|---|---|---|---|---|---|---|---|---|---|
| | | 11 | 21 | 31 | 12 | 22 | 32 | 13 | 23 | 33 |
| i | 1 | 0 | 0 | $\chi_{131}$ | 0 | 0 | 0 | $\chi_{113}$ | 0 | 0 |
| | 2 | 0 | 0 | 0 | 0 | 0 | $\chi_{232}$ | 0 | $\chi_{223}$ | 0 |
| | 3 | $\chi_{311}$ | 0 | 0 | 0 | $\chi_{322}$ | 0 | 0 | 0 | $\chi_{333}$ |

Number of independent coefficients: 4

$\chi_{131} = \chi_{232}, \; \chi_{113} = \chi_{223}, \; \chi_{311} = \chi_{322}$

### -6

| $\chi_{ijk}$ | | \multicolumn{9}{c}{jk} | | | | | | | | |
|---|---|---|---|---|---|---|---|---|---|---|
| | | 11 | 21 | 31 | 12 | 22 | 32 | 13 | 23 | 33 |
| i | 1 | $\chi_{111}$ | $\chi_{121}$ | 0 | $\chi_{112}$ | $\chi_{122}$ | 0 | 0 | 0 | 0 |
| | 2 | $\chi_{211}$ | $\chi_{221}$ | 0 | $\chi_{212}$ | $\chi_{222}$ | 0 | 0 | 0 | 0 |
| | 3 | 0 | 0 | 0 | 0 | 0 | 0 | 0 | 0 | 0 |

Number of independent coefficients: 2

$\chi_{111} = -\chi_{122} = -\chi_{221} = -\chi_{212}, \; \chi_{112} = \chi_{121} = \chi_{221} = -\chi_{222}$

### -6m2

| $\chi_{ijk}$ | | \multicolumn{9}{c}{jk} | | | | | | | | |
|---|---|---|---|---|---|---|---|---|---|---|
| | | 11 | 21 | 31 | 12 | 22 | 32 | 13 | 23 | 33 |
| i | 1 | 0 | $\chi_{121}$ | 0 | $\chi_{112}$ | 0 | 0 | 0 | 0 | 0 |
| | 2 | $\chi_{211}$ | 0 | 0 | 0 | $\chi_{222}$ | 0 | 0 | 0 | 0 |
| | 3 | 0 | 0 | 0 | 0 | 0 | 0 | 0 | 0 | 0 |

Number of independent coefficients: 1

$\chi_{112} = \chi_{121} = \chi_{211} = -\chi_{222}$

### 6mm

| $\chi_{ijk}$ | | \multicolumn{9}{c}{jk} | | | | | | | | |
|---|---|---|---|---|---|---|---|---|---|---|
| | | 11 | 21 | 31 | 12 | 22 | 32 | 13 | 23 | 33 |
| i | 1 | 0 | 0 | $\chi_{131}$ | 0 | 0 | 0 | $\chi_{113}$ | 0 | 0 |
| | 2 | 0 | 0 | 0 | 0 | 0 | $\chi_{232}$ | 0 | $\chi_{223}$ | 0 |
| | 3 | $\chi_{311}$ | 0 | 0 | 0 | $\chi_{322}$ | 0 | 0 | 0 | $\chi_{333}$ |

Number of independent coefficients: 4

$\chi_{131} = \chi_{232}, \; \chi_{113} = \chi_{223}, \; \chi_{311} = \chi_{322}$

## §2 Sample preparation

We provide the details of the single crystal growth used in our study. Single crystals of Te were produced via the physical vapour transport (PVT) method[5]. A mass of 0.36 g of bulk Te was vacuum-sealed in a quartz tube and heated in a three-zone electric furnace. The temperatures at the source side, $T_1$, and on the crystal growth side, $T_2$ and $T_3$, were adjusted following the time profile shown in Fig. S1a. A picture of the obtained crystal is shown in Fig. S1b. The typical size was ~1 cm in the direction of the *c*-axis, with a thickness of 2–4 mm.

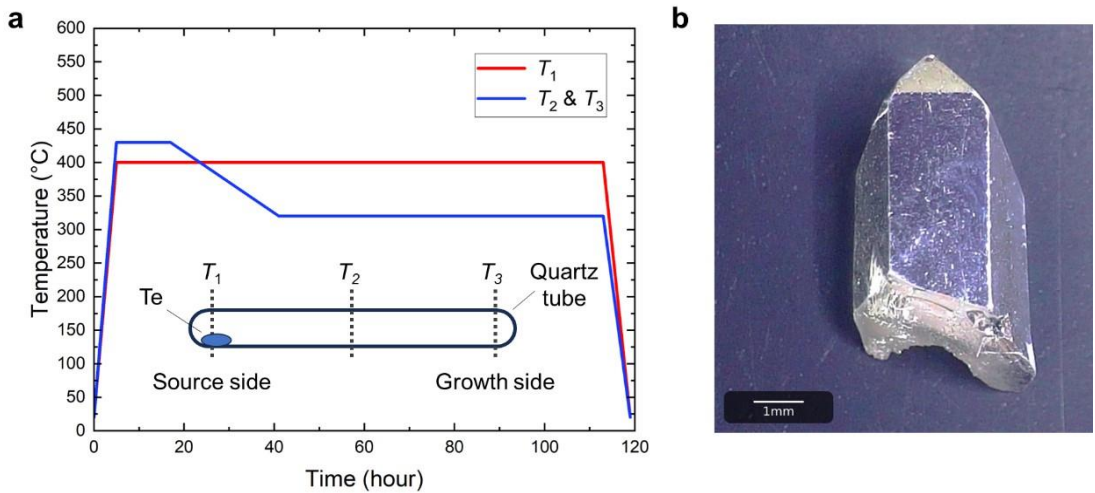

**Figure S1 | Synthesis of the Te crystals. a**, Time profile of Te synthesis via the PVT method. **b**, Obtained crystal of Te.

## §3 Electrical properties of the tellurium samples

Fig. S2 shows the temperature dependence of the electrical resistivity for samples A and B with different chiralities. As shown in the figure, the temperature dependences of the resistivities of samples A and B are very similar, and their absolute values are close. Therefore, the possibility of a difference in the temperature dependence of the electrical resistivity as a reason for the sign reversal of the NCTE Hall effect is ruled out. The behaviour at high temperatures ($T > 200$ K) is quintessentially semiconducting, whereas the medium temperature region (50 K $< T <$ 200 K) shows a flat temperature dependence due to the contribution of the impurity carriers. At low temperatures ($T < 50$ K), the resistivity increases due to the freezing of the carriers and shows a typical behaviour of variable range hopping.

Fig. S3a and b shows the normal Hall effects for samples A and B measured at 300 K. The Hall coefficient is obtained by linear fitting of the experimental data up to 1 T, with $R_H = -6.88 \times 10^{-5}$ m$^3$C$^{-1}$ for sample A and $R_H = -1.30 \times 10^{-4}$ m$^3$C$^{-1}$ for sample B. Consequently, the carrier density is $n = 9.07 \times 10^{22}$ m$^{-3}$ for sample A and $n = 4.78 \times 10^{22}$ m$^{-3}$ for sample B. These values are nearly the same as the typical values reported for the Te samples[6]. Since both samples exhibit a negative Hall coefficient, we can conclude that electrons are the dominant carriers near room temperature. The deviation from linearity at high magnetic fields is likely caused by the effect of mixed contributions from multiple carriers.

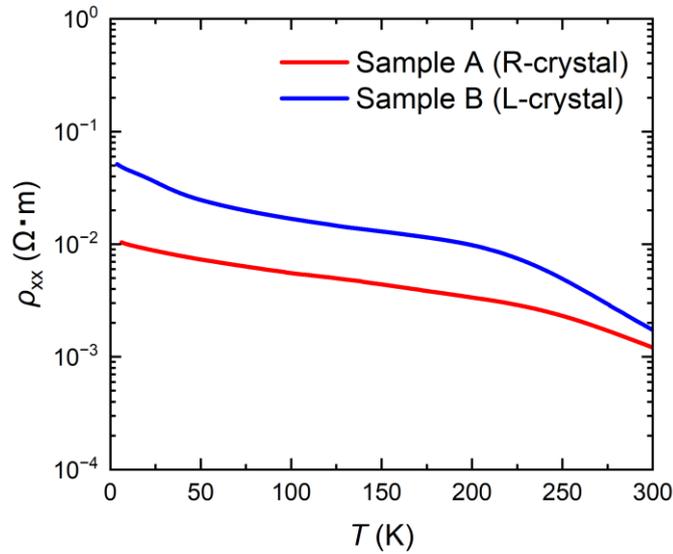

**Figure S2 | Temperature dependence of the resistivity of tellurium.** The red and blue solid lines correspond to samples A (right-handed crystal) and B (left-handed crystal), respectively.

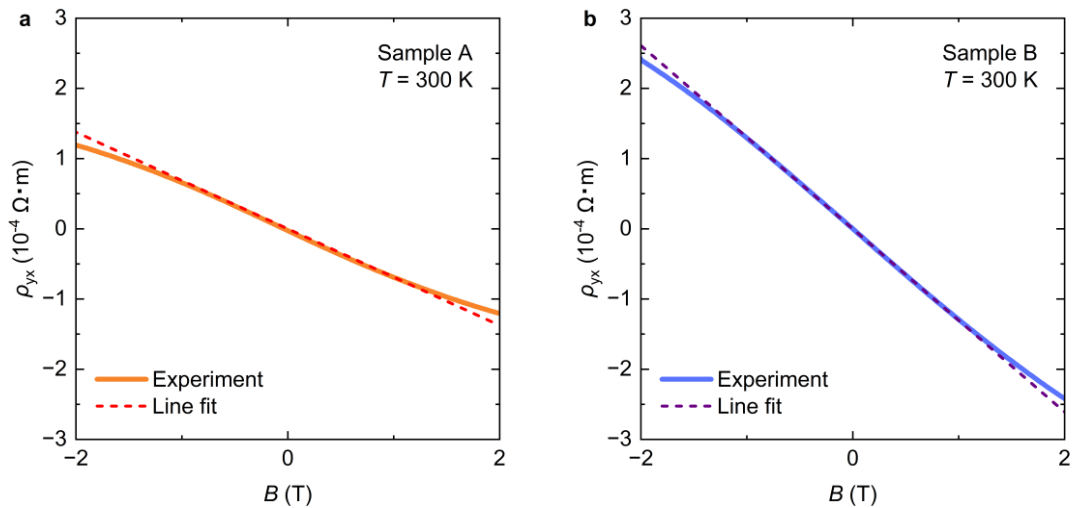

**Figure S3 | Normal Hall resistivity at 300 K. a**, Hall resistivity of sample A. The orange solid line and red dashed line represent the experimental data and the linear fitting curve for ±1 T data,

respectively. **b**, Hall resistivity of sample B. The blue solid line and purple dashed line represent the experimental data and the linear fitting curve for ±1 T data, respectively.

## §4 Microscopic theory of the NCTE Hall effect

According to microscopic theory[7], the total NCTE Hall conductivity $\sigma_{\text{NCTE}}$ can be described as the sum of contributions from the BCD, $\sigma_z^{\text{BCD}}$, and orbital magnetization, $\sigma_z^{\text{OM}}$; these are defined as follows:

$$\sigma_z^{\text{BCD}} = e^2\tau \sum_{n,k}(\varepsilon_{nk} - \mu)\left(-\frac{\partial f}{\partial \varepsilon}\right)_{\varepsilon=\varepsilon_{nk}} \times \left[(\partial_{k_z}\varepsilon_{nk})\Omega_n^z - \frac{1}{2}\left\{(\partial_{k_x}\varepsilon_{nk})\Omega_n^x + (\partial_{k_y}\varepsilon_{nk})\Omega_n^y\right\}\right], \tag{S9}$$

$$\sigma_z^{\text{OM}} = -e\tau \sum_{n,k}(\varepsilon_{nk} - \mu)\left(-\frac{\partial f}{\partial \varepsilon}\right)_{\varepsilon=\varepsilon_{nk}} \boldsymbol{\nabla}_k \cdot \boldsymbol{m}_{nk}^\perp, \tag{S10}$$

where $f$ and $\boldsymbol{\Omega}_n = \boldsymbol{\nabla}_k \times \boldsymbol{A}_n(\boldsymbol{k})$ are the Fermi distribution function and the Berry curvature with $\boldsymbol{A}_n(\boldsymbol{k}) = -i\langle n(\boldsymbol{k})|\boldsymbol{\nabla}_k n(\boldsymbol{k})\rangle$ as the Berry connection, respectively; and $\varepsilon_{nk}$ and $|n(\boldsymbol{k})\rangle$ are the eigenenergy and eigenvector of the Hamiltonian. In addition, $\boldsymbol{m}_{nk}^\perp = \boldsymbol{m}_{nk} - m_{nk}^z \hat{\boldsymbol{e}}_z = (m_{nk}^x, m_{nk}^y, 0)$ is the orbital magnetic moment written by the interband Berry connection,

$$\boldsymbol{m}_{nk} = \frac{e}{2}\sum_m (\varepsilon_{mk} - \varepsilon_{nk})\,\text{Im}[\boldsymbol{A}_{nm}(\boldsymbol{k}) \times \boldsymbol{A}_{mn}(\boldsymbol{k})], \tag{S11}$$

$$\boldsymbol{A}_{mn}(\boldsymbol{k}) = -i\langle n(\boldsymbol{k})|\boldsymbol{\nabla}_k m(\boldsymbol{k})\rangle. \tag{S12}$$

An important feature is that both contributions are proportional to the square of the electron charge $e^2$. Thus, the sign of the NCTE Hall effect is independent of the type of carrier. The band structure near the top of the valence band is approximately isotropic in Te, which is expected to suppress the contribution of the Berry curvature dipole $\sigma_z^{\text{BCD}}$. Therefore, the contribution from the orbital magnetic moment $\sigma_z^{\text{OM}}$, which is predicted from microscopic theory, is very important[3,7]. $\sigma_z^{\text{OM}}$ can be interpreted as the Nernst effect due to orbital magnetisation caused by the applied current or the Hall effect due to orbital magnetisation caused by the temperature gradients.